\documentclass{article}

\usepackage[utf8]{inputenc}
\usepackage[T1]{fontenc}
\usepackage[margin=2.5cm]{geometry}
\usepackage{graphicx}
\usepackage{dcolumn}
\usepackage{bm}
\usepackage{amsmath}
\usepackage{amssymb}
\usepackage{xcolor}
\usepackage{authblk}
\usepackage[numbers,sort&compress]{natbib}
\usepackage[colorlinks=true,citecolor=blue,linkcolor=red,urlcolor=blue]{hyperref}

\makeatletter
\newcommand{\blfootnote}[1]{%
  \begingroup
  \renewcommand{\thefootnote}{}\footnote{#1}%
  \addtocounter{footnote}{-1}%
  \endgroup
}
\makeatother

\title{Heat Capacity Anomalies and Thermal Crossovers in Finite
Su-Schrieffer-Heeger Chains}

\author[1]{Carlos Magno da Concei\c{c}\~ao\thanks{carlosmagno@id.uff.br}}
\author[2]{Julio C\'esar P\'erez-Pedraza\thanks{julio.perez@correo.nucleares.unam.mx}}
\author[3,5]{Alfredo Raya\thanks{alfredo.raya@umich.mx}}
\author[4,5]{\\ Cristi\'an Villavicencio\thanks{cvillavicencio@ubiobio.cl (corresponding author)}}

\affil[1]{Universidade Federal Fluminense -- RHS/RCN, 28895-532 Rio das Ostras, Rio de Janeiro, Brazil}
\affil[2]{Instituto de Ciencias Nucleares, Universidad Nacional Aut\'onoma de M\'exico, Circuito Exterior s/n, Ciudad Universitaria, Col. Universidad Nacional Aut\'onoma de M\'exico, Apartado Postal 70-543, C.P. 04510, Ciudad de M\'exico, Mexico}
\affil[3]{Faculty of Electric Engineering, Universidad Michoacana de San Nicol\'as de Hidalgo, Edificio $\Omega$, Ciudad Universitaria, Morelia, 58040, Mexico}
\affil[4]{Departamento de Ciencias B\'asicas de la Facultad de Ciencias, Universidad del B\'io-B\'io, Casilla 447, Chill\'an, Chile}
\affil[5]{Centro de Ciencias Exactas, Universidad del B\'io-B\'io, Casilla 447, Chill\'an, Chile}

\date{}
\begin{document}

\maketitle
\blfootnote{The authors are listed in alphabetical order.}

\begin{abstract}
We investigate the thermodynamic properties of finite Su-Schrieffer-Heeger (SSH) chains in thermal equilibrium at fixed temperature and chemical potential. Using the canonical and grand canonical ensembles, we calculate the energy density, particle number density, entropy, and heat capacity as functions of temperature, chemical potential, and hopping asymmetry. Our analysis reveals a double-peak anomaly in the heat capacity for non-dimerized configurations, in which two maxima are separated by a local minimum. We identify this feature as a Schottky-type thermal crossover governed by two distinct energy scales. %rather than a genuine phase transition, since the free energy of a finite chain is an analytic function of temperature and cannot support a thermodynamic singularity.
The anomaly is most pronounced when the chemical potential is comparable to the band scale, and its two maxima separate further as the hopping asymmetry increases and the chain length grows, reflecting the widening gap between the two energy scales and the increasing density of states.
We demonstrate that while the topological properties are determined by boundary states, the bulk thermodynamic behavior exhibits a rich crossover structure that can be tuned through the hopping parameter ratio. These findings provide insights into the interplay between topology, finite-size effects, and thermal fluctuations in one-dimensional topological systems, with potential implications for experimental realizations in cold atoms, photonic systems, and topoelectrical circuits.
\end{abstract}

\section{Introduction}

The discovery of topological phases of matter has fundamentally reshaped our understanding of quantum materials, revealing that global topological invariants can protect physical properties against local perturbations~\cite{Klitzing1980,Thouless1982,Haldane1988}. Topological insulators, which exhibit insulating behavior in the bulk while hosting conducting states at their boundaries, have emerged as a paradigm for understanding the deep connection between topology and condensed matter physics~\cite{Kane2005,HasanKane2010,QiZhang2011}. The robustness of these edge states, protected by time-reversal or other fundamental symmetries, has opened avenues for applications in quantum computing, spintronics, and dissipationless electronics, among other important fields.

From a plethora of models describing topological insulators, one-dimensional (1D) systems hold particular significance due to their analytical tractability and experimental accessibility. The Su-Schrieffer-Heeger (SSH) model~\cite{SSH1979}, originally introduced to describe solitons in polyacetylene, has become the quintessential example of a 1D topological insulator. Despite its simplicity—consisting of a dimerized chain with alternating hopping amplitudes—the SSH model captures essential features of topological phases, including quantized Berry phases, bulk-boundary correspondence, and topologically protected edge states~\cite{Asboth2016,Shen2017,BernevigHughes2013}. The elegance of the model lies in its ability to exhibit a topological phase transition by simply tuning the ratio of intra-cell to inter-cell hopping parameters, demonstrating that topology can emerge from symmetry breaking in lattice systems.

In recent years, the community has witnessed experimental realizations of the SSH model beyond traditional condensed matter systems. Photonic lattices~\cite{Li2014SSH}, ultracold atomic gases~\cite{Atala2013}, mechanical metamaterials~\cite{Susstrunk2015}, and topoelectrical circuits~\cite{Ningyuan2015,Lee2018,PerezPedraza2024,ANZUREZ2025417609} have all successfully implemented SSH physics, allowing for unprecedented control over system parameters and direct observation of topological phenomena. These platforms have not only confirmed theoretical predictions but have also revealed new aspects of topological protection under dissipation, non-Hermiticity~\cite{YaoWang2018}, and long-range interactions~\cite{Li2014SSH}. Moreover, the recent realization of 1D topological insulators in ultrathin germanene nanoribbons~\cite{Klaassen2025} with persistence of edge states down to $\sim 2$ nm width demonstrates the relevance of finite-size effects and opens possibilities for nanoscale topological devices.

While the topological properties of the SSH model have been extensively studied at zero temperature, understanding its thermodynamic behavior presents distinct challenges and opportunities. Thermodynamic properties encode information about the density of states, level statistics, and collective excitations that complement topological characterization.
In finite-size systems, thermal fluctuations can compete with boundary effects and produce heat capacity anomalies and crossovers that are invisible to the topological classification. We must stress that these are finite-size features and not new thermodynamic phases: topological invariants such as the Zak phase are defined for translationally invariant systems in the thermodynamic limit, and the presence of boundary states in a finite chain does not by itself certify topological order.
Furthermore, realistic implementations of topological systems operate at finite temperatures, making thermodynamic analysis crucial for practical applications.

The thermodynamics of topological systems has attracted growing attention in recent years. Studies have explored thermal transport~\cite{Sheng2006}, thermoelectric properties~\cite{Takahashi2007}, thermodynamic signatures in topological phases~\cite{Kempkes2016UniversalitiesThermoTopo}, topological invariants and transitions at nonzero temperature~\cite{MoligniniCooper2023}, and thermodynamic and spectral properties in SSH/Peierls-type chains~\cite{Weber2016AdiabaticPeierls}. For non-Hermitian generalizations of the SSH model, thermodynamic considerations have revealed unconventional phase transitions and entanglement properties~\cite{MunozArboleda2024}. In related SSH-like settings, grand canonical Peierls physics has also revealed metastable behavior~\cite{Jeckelmann2015GrandCanonicalPeierls}. However, a comprehensive analysis of the thermodynamics of the Hermitian SSH model, particularly focusing on finite-size effects and the role of chemical potential in systems with open boundary conditions, remains largely unexplored.

Thermodynamic response functions in finite-size systems present unique features compared to the thermodynamic limit.
A genuine phase transition requires a non-analyticity of the free energy, which can only occur in the above mentioned limit; in a finite system the partition function is a finite sum of exponentials and the free energy stays analytic in temperature.
What survives at finite size are rounded anomalies in response functions such as heat capacity and magnetic susceptibility~\cite{Binder1987,Privman1990},
together with crossovers between regimes controlled by different energy scales.
Finite-size scaling theory~\cite{Fisher1972} describes how such rounded features approach true critical behavior as the system grows, but only in models that host a transition in the first place. For a one-dimensional chain of non-interacting fermions with short-range hopping, no finite-temperature transition exists in any limit, so the anomalies we report below are understood as crossovers rather than precursors of criticality.

We investigate the thermodynamic properties of finite SSH chains using the grand canonical ensemble formalism, allowing for particle exchange with a reservoir at fixed chemical potential and temperature. We consider a generalized parametrization of the hopping amplitudes that interpolates between symmetric ($v=w$) and fully dimerized ($v=0$ or $w=0$) configurations. Our main findings include:
(i) a double-peak anomaly in the heat capacity of non-dimerized chains, consisting of two maxima separated by a local minimum, which we identify as a Schottky-type thermal crossover between two energy scales rather than a phase transition;
(ii) the two maxima separate further and the intervening minimum deepens as the hopping asymmetry increases and as the chain length grows, a behavior that reflects the widening separation of the two energy scales and the growing density of states, not an emergent criticality;
(iii) this crossover is distinct from the topological phase transition: while topology is determined by edge states and the winding number, the thermodynamic anomalies reflect bulk properties encoded in the density of states;
(iv) for closed systems, i.e.,  with fixed particle number with energy exchange only, the anomaly persists but with reduced magnitude, indicating that particle fluctuations enhance the thermodynamic signatures;
(v) the chemical potential exhibits distinct temperature dependence regimes, with strong sensitivity to hopping asymmetry at low temperatures ($k_BT < \epsilon$) and universal behavior at high temperatures.
These results demonstrate that finite SSH chains possess rich thermodynamic structure beyond their topological classification. The observed heat capacity anomalies provide experimentally accessible signatures that could be measured in quantum simulator platforms such as cold atoms or topoelectrical circuits. Moreover, our findings suggest that tuning the hopping asymmetry offers a control parameter for engineering the thermodynamic response and the position of the crossover, potentially useful for thermal management in nanoscale topological devices.

For the presentation of our results, the remainder of this article is organized as follows: In Sec.~\ref{sec:model}, we present the SSH model with open boundary conditions and derive the energy spectrum for finite chains. Section~\ref{sec:thermo} develops the thermodynamic formalism within the grand canonical ensemble, deriving expressions for the grand canonical potential, energy, particle number, entropy, and heat capacity. In Sec.~\ref{sec:results}, we present our numerical results for two scenarios: fixed chain length with open boundary conditions and fixed particle number with variable chain length. We analyze the heat capacity, energy density, and chemical potential as functions of temperature and hopping asymmetry, identifying the double-peak anomaly and discussing its properties. Finally, Sec.~\ref{sec:conclusions} summarizes our findings and discusses potential extensions and experimental implications of this work.

\section{The Model}
\label{sec:model}

The Su-Schrieffer-Heeger model describes a one-dimensional chain of atoms arranged in a bipartite lattice with sublattices $A$ and $B$. Within the tight-binding approximation, electrons can hop between nearest-neighbor sites with amplitudes that alternate between two values, leading to a dimerization of the lattice. This dimerization breaks the translational symmetry and is responsible for the topological properties of the system.

\subsection{Hamiltonian and Wannier states}

\begin{figure}
    \centering
    \includegraphics[scale=0.7]{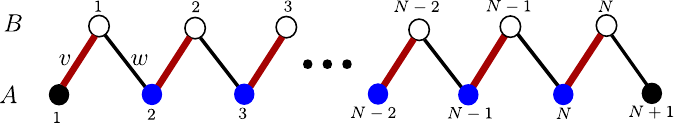}
    \caption{Illustration of the crystal structure in our model of a fully dimerized finite chain of atoms with $N+1$ sites. Hopping parameters are shown as thick and thin lines.}
    \label{fig:model}
\end{figure}

We consider a finite SSH chain with $N$ unit cells, each containing one $A$ site and one $B$ site, as illustrated in Fig.~\ref{fig:model}. The system is described by localized Wannier states $|n,X\rangle$, where $n=1,2,\ldots,N$ labels the unit cell and $X\in\{A,B\}$ denotes the sublattice. The Hamiltonian for the SSH model with nearest-neighbor hopping is:

\begin{equation}
H = \sum_{n=1}^{N} \left\{ v \left( |n,A\rangle\langle n,B| + |n,B\rangle\langle n,A| \right) + w \left( |n+1,A\rangle\langle n,B| + |n,B\rangle\langle n+1,A| \right) \right\},
\label{eq:hamiltonian}
\end{equation}
\noindent
where $v$ represents the intra-cell hopping amplitude (between $A$ and $B$ sites within the same unit cell) and $w$ is the inter-cell hopping amplitude (between adjacent unit cells). The relative strength of these hopping parameters determines the topological phase of the system: when $|v| > |w|$, the chain is in the trivial phase, while $|v| < |w|$ corresponds to the topological phase with protected edge states, as discussed in the textbooks~\cite{Asboth2016,Shen2017,BernevigHughes2013}.

\subsection{Energy eigenvalues and eigenstates}

To find the energy eigenvalues and eigenstates, inspired by the rationale on Refs.~\cite{Magno}, and following the closed-form treatments of tridiagonal SSH-type Hamiltonians developed earlier by Sirker \textit{et al.}~\cite{SirkerBoundary2014}, Shin~\cite{Shin1997Tridiagonal}, and Li and Miroshnichenko~\cite{LiMiroshnichenko2019}, we expand a general state in the system in the form
\begin{equation}
|\phi\rangle = \sum_{n=1}^{N} \left( a_n |n,A\rangle + b_n |n,B\rangle \right),
\label{eq:eigenstate}
\end{equation}
\noindent
where $a_n = \langle n,A|\phi\rangle$ and $b_n = \langle n,B|\phi\rangle$ are the probability amplitudes for finding the electron in sublattice $A$ or $B$ at cell $n$, respectively. The time-independent Schr\"odinger equation $H|\phi\rangle = E|\phi\rangle$ yields the coupled system of equations:

\begin{align}
v b_n + w b_{n-1} &= E a_n, \label{eq:schrodinger_a} \\
v a_n + w a_{n+1} &= E b_n. \label{eq:schrodinger_b}
\end{align}
\noindent
Eliminating $b_n$ from these equations leads to a recursion relation for the amplitudes $a_n$:
\begin{equation}
a_{n+1} = p\, a_n + q\, a_{n-1},
\label{eq:recursion}
\end{equation}
\noindent
with coefficients
\begin{equation}
p = \frac{E^2 - v^2 - w^2}{vw}, \quad q = -1.
\label{eq:coefficients}
\end{equation}

\subsection{Boundary conditions and energy spectrum}

For a finite chain with open boundary conditions, we impose that the wavefunction vanishes at the boundaries: $a_1 = a_{N+1} = 0$. This constraint physically represents either a chain with hard-wall boundaries or an impurity that causes the wavefunction to collapse at the endpoints, naturally introducing an asymmetry between the sublattices $A$ and $B$.
We emphasize that this hard-wall condition defines a well-posed but specific boundary problem that is not identical to the standard open SSH chain obtained by cutting the terminal bonds. The condition $a_1 = a_{N+1} = 0$ forces the amplitudes to vanish at both ends and, in doing so, suppresses the near-zero-energy edge states that characterize the topologically non-trivial phase of the standard open chain. The spectrum in Eq.~(\ref{eq:energy_spectrum}) therefore describes this hard-wall SSH chain rather than the conventional open SSH chain, and the discrete label $m$ enumerates bulk-like standing waves. Because our thermodynamic analysis probes bulk properties encoded in the full density of states, the conclusions below do not rely on the presence of edge states.
The recursion relation in Eq.~(\ref{eq:recursion}) has the general solution:
\begin{equation}
a_n = A_+ \lambda_+^n + A_- \lambda_-^n,
\label{eq:general_solution}
\end{equation}
\noindent
where
\begin{equation}
\lambda_\pm = \frac{p \pm \sqrt{p^2 + 4q}}{2}.
\label{eq:lambda}
\end{equation}

Using the Binet formula, we can express $a_n$ in terms of generalized Fibonacci polynomials $F_n(p,q)$ and generalized Lucas polynomials $L_n(p,q)$:
\begin{equation}
a_n = F_{n-1}(p,q) a_2 + [L_{n-1}(p,q) - F_n(p,q)] a_1,
\label{eq:fibonacci_solution}
\end{equation}
\noindent
where
\begin{equation}
F_n(p,q) = \frac{\lambda_+^n - \lambda_-^n}{\lambda_+ - \lambda_-}, \quad L_n(p,q) = \lambda_+^n + \lambda_-^n.
\label{eq:fibonacci_lucas}
\end{equation}

The boundary condition $a_1 = a_{N+1} = 0$ requires that $F_N(p,q) = 0$, which from the Binet form gives:
\begin{equation}
\frac{\lambda_+^N - \lambda_-^N}{\lambda_+ - \lambda_-} = 0 \quad \Rightarrow \quad \left(\frac{\lambda_+}{\lambda_-}\right)^N = 1.
\label{eq:boundary_quantization}
\end{equation}
\noindent
Thus, this quantization condition is satisfied when $(\lambda_+/\lambda_-)^N = e^{2\pi i m}$ for integer $m$, leading to quantized values of the parameter $p$ and thus discrete energy levels.
We have checked that the closed-form spectrum obtained this way coincides, level by level, with a direct numerical diagonalization of the finite Hamiltonian in Eq.~(\ref{eq:hamiltonian}) subject to the same boundary condition, so the analytic expression and exact diagonalization are used here as mutually consistent routes to the spectrum.

\subsection{Parametrization and energy spectrum}

To explore the phase diagram systematically, we introduce a geometrical parametrization of the hopping amplitudes in terms of the angle $\theta$ as follows:
\begin{equation}
v = \epsilon \cos(\theta + \pi/4), \quad w = \epsilon \sin(\theta + \pi/4),
\label{eq:parametrization}
\end{equation}
\noindent
such that $\epsilon = \sqrt{v^2 + w^2}$ sets the overall energy scale and $\theta \in [-\pi/4, \pi/4]$ parameterizes the hopping asymmetry. This parametrization has the property that $\theta = 0$ corresponds to the symmetric case $v = w$, while $\theta = \pm \pi/4$ gives the fully dimerized limits $v = 0$ (at $\theta = +\pi/4$) or $w = 0$ (at $\theta = -\pi/4$).
The ratio $v/w = \cot(\theta + \pi/4)$ provides a direct measure of the asymmetry.

With this parametrization and considering the quantization from boundary conditions, the energy spectrum for a finite SSH chain becomes:
\begin{equation}
E_m = \pm\, \epsilon \sqrt{1 + \cos(2\theta) \cos\left(\frac{m\pi}{N}\right)},
\label{eq:energy_spectrum}
\end{equation}
\noindent
where $\pm $ labels the upper ($+$) and lower ($-$) bands, and $m=1,2,...,N-1$ is the quantum number arising from the boundary conditions. The term $\cos(2\theta)$ inside the square root encodes the band gap modulation due to dimerization. It is crucial to specify the state counting under the hard-wall constraints: the condition $a_1=a_{N+1}=0$ strictly projects out the boundary degrees of freedom, removing the two near-zero-energy edge modes characteristic of the standard open topological chain. Consequently, the discrete label $m$ runs over exactly $N-1$ bulk standing waves for each band. Summing over the particle-hole symmetric bands and accounting for the physical spin degeneracy, the total number of accessible single-particle states is properly conserved for the truncated bulk, ensuring that the subsequent thermodynamic trace avoids double counting and correctly normalizes the intensive properties.

Several important features can be straightforwardly extracted from Eq.~(\ref{eq:energy_spectrum}).  To start with, for $\theta = 0$ (symmetric hopping, $v=w$), the spectrum simplifies and the gap closes at the band center. For $\theta = \pm\pi/4$ (fully dimerized), the spectrum exhibits maximum asymmetry between bands. Moreover, the spectrum is symmetric under $\theta \to -\theta$ due to the equivalence of the two dimerization patterns in a finite chain with collapsed wavefunction at the boundaries.
This energy spectrum forms the basis for our thermodynamic analysis in the following section. Note that while the topological phase transition occurs at $|v| = |w|$ or, equivalently, $\theta = 0$, the thermodynamic properties depend continuously on $\theta$ and exhibit their own distinct crossover structure.

\begin{figure}
    \centering
\includegraphics[width=0.5\linewidth]{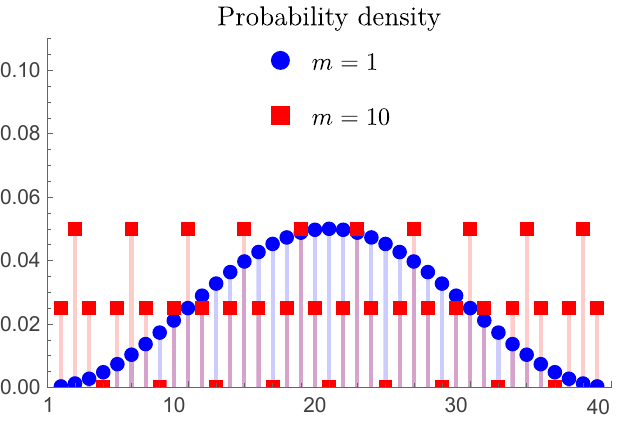}
    \caption{Real-space probability density $|\psi|^2$ of the $m=1$ and $m=10$ levels, for $N=20$. Here, sites 1 and 41 reflect correctly the condition $a_1=a_{41}=0$.}
    \label{fig:rho_1_10}
\end{figure}

Consistent with the discussion of the hard-wall boundary condition above, the lowest-lying level ($m=1$) is the lowest bulk standing wave of this problem and is not an edge-localized mode; the hard-wall condition removes the near-zero-energy edge states of the standard open chain. Higher $m$ values correspond to bulk states of shorter wavelength. To explicitly demonstrate the bulk character of these solutions, Fig. \ref{fig:rho_1_10} shows the real-space probability density $\vert{}\psi_n^{(m)}\vert{}^2$ for the $m=1$ and $m=10$ levels, for $N=20$. As expected from the imposed boundary conditions, the wavefunctions are strictly confined and vanish at the ends of the chain. The lowest-energy state ($m=1$) forms a perfect half-sine bulk standing wave rather than a topologically protected boundary state. Consequently, our thermodynamic analysis intrinsically probes the bulk properties of the finite chain rather than boundary phenomena.

\section{Thermodynamics}
\label{sec:thermo}

In this section, we develop the thermodynamic description of the finite SSH chain using the grand canonical ensemble considering  a thermal reservoir, characterized by a temperature $T$ and a chemical potential $\mu$.

\subsection{Grand canonical ensemble and partition function}

The grand canonical partition function for the SSH chain is defined as:
\begin{equation}
Z = \text{Tr}\, e^{-\beta(H - \mu_A Q_A - \mu_B Q_B)},
\label{eq:partition_function}
\end{equation}
\noindent
where $\beta = 1/(k_B T)$ is the inverse temperature, 
\begin{equation}
Q_X = \sum_{n=1}^N |n,X\rangle\langle n,X|,
\end{equation}
are the occupation number operators for each sublattice, and $\mu_X$ are the corresponding chemical potentials. The inclusion of distinct chemical potentials for the two sublattices allows us to study asymmetric charge distributions, although in this work we focus on the case $\mu_A = \mu_B = \mu$ to ensure overall charge conservation.

Since fermions obey Pauli exclusion principle, each energy level can be occupied by at most one electron (per spin). Including spin degeneracy (factor of 2), and treating particles and holes on equal footing, the partition function can be evaluated explicitly. The grand canonical potential is:

\begin{equation}
\Omega(T,\mu) = -k_B T \ln Z = -\frac{2}{\beta}  \sum_{m=1}^{N-1} \left\{ \ln\left[1 + e^{-\beta(E_m - \mu)}\right] + \ln\left[1 + e^{-\beta(E_m + \mu)}\right] \right\},
\label{eq:grand_potential}
\end{equation}
\noindent
where the factor 2 accounts for spin degeneracy, and the two logarithmic terms correspond to particles (with energy $E_m - \mu$) and holes (with energy $E_m + \mu$), respectively. %\revnote{With the chemical potential now residing only in these Fermi weights, the energy and particle-number expressions below are free of the double counting present in the first version.} \revnote{Since the spectrum $E_m^s$ is already particle--hole symmetric under $s\to -s$, care must be taken that the particle and hole terms together with the sum over $s=\pm$ count each physical single-particle state exactly once; state this counting explicitly so that $\bar N$ and $\bar E$ reduce to the standard free-fermion sums.}

\subsection{Thermodynamic quantities}

From the grand canonical potential, all equilibrium thermodynamic properties can be derived. For instance, the entropy and mean particle number are obtained through standard thermodynamic relations:
\begin{align}
S &= -\left(\frac{\partial \Omega}{\partial T}\right)_\mu, \label{eq:entropy} \\
\bar N &= -\left(\frac{\partial \Omega}{\partial \mu}\right)_T. \label{eq:particle_number}
\end{align}
Note that whereas $N$ denotes the number of unit cells of the chain (equivalently the number of sites of each sublattice), $\bar N$ denotes the mean particle number.

Explicitly, for the mean energy and mean particle number, we have

\begin{align}
\bar{E} &= \sum_{m=1}^{N-1} E_m^s \left[ n_F(E_m - \mu) + n_F(E_m + \mu) \right], \label{eq:energy} \\
\bar{N} &=  \sum_{m=1}^{N-1} \left[ n_F(E_m- \mu) - n_F(E_m + \mu) \right], \label{eq:number}
\end{align}
\noindent
where $n_F(E) = 1/[1 + e^{\beta E}]$ is the Fermi-Dirac distribution function. The mean energy $\bar{E}$ is related to the grand canonical potential by $\bar{E} = \Omega + TS + \mu\bar{N}$.

\subsection{Heat capacity}

The heat capacity is a central quantity in thermodynamic phase transition theory, as it exhibits characteristic signatures such as peaks, discontinuities, or divergences at critical points. For our system, we must carefully distinguish between different statistical ensembles.

In the canonical ensemble (fixed particle number $\bar{N}$, variable $\mu$), the heat capacity is:
\begin{equation}
C_{\bar{N}} = \left(\frac{\partial \bar{E}}{\partial T}\right)_{\bar{N}}.
\label{eq:heat_capacity_definition}
\end{equation}

Since both $\bar{E}$ and $\bar{N}$ depend on $T$ and $\mu$, we must account for the implicit temperature dependence of the chemical potential. Using the chain rule:

\begin{equation}
C_{\bar{N}} = \left(\frac{\partial \bar{E}}{\partial T}\right)_\mu + \left(\frac{\partial \bar{E}}{\partial \mu}\right)_T \left(\frac{\partial \mu}{\partial T}\right)_{\bar{N}} = \left(\frac{\partial \bar{E}}{\partial T}\right)_\mu - \left(\frac{\partial \bar{E}}{\partial \mu}\right)_T \frac{\left(\frac{\partial \bar{N}}{\partial T}\right)_\mu}{\left(\frac{\partial \bar{N}}{\partial \mu}\right)_T}.
\label{eq:heat_capacity}
\end{equation}
\noindent
In the second equality, we used the implicit function theorem to express $(\partial\mu/\partial T)_{\bar{N}}$ in terms of derivatives of $\bar{N}(T,\mu)$.
The heat capacity per site, $C_{\bar{N}}/N$, provides a size-independent measure of thermal response. At second-order phase transitions in finite systems, $C_{\bar{N}}$ typically exhibits rounded peaks or shoulders whose height and width scale with system size according to finite-size scaling theory~\cite{Binder1987}.

\subsection{Dimensionless quantities and parameter space}

To facilitate numerical analysis and identify universal features, we introduce dimensionless quantities by scaling with the energy parameter $\epsilon$:
\begin{align}
\tilde{T} &= k_B T/\epsilon, \qquad \tilde{\mu} = \mu/\epsilon, \\
\tilde{E} &= \bar{E}/\epsilon, \qquad \tilde{C} = C_{\bar N}/k_B.
\end{align}
Since $\epsilon = \sqrt{v^2 + w^2}$ is the only energy scale in the single-particle problem, we set $\epsilon = 1$ throughout and measure every energy, temperature, and chemical potential in units of $\epsilon$. This is what makes the scaling meaningful: all curves at different hopping asymmetries share the same energy unit, and the value of $\epsilon$ never needs to be specified separately.
The thermodynamic behavior of the system is then characterized by the dimensionless parameters: $\tilde{T}$, $\tilde{\mu}$, $N$ (chain length in unit cells), $\bar{N}$ (particle number), and $\theta$ (hopping asymmetry parameter). Recall that the topological phase transition occurs at $\theta = 0$, providing a natural reference point in parameter space.

In the following section, we present numerical results exploring two complementary scenarios, the grand canonical and the canonical ensembles.  These scenarios correspond to different experimental situations and reveal distinct aspects of the thermodynamic phase structure.

\section{Results and Discussion}
\label{sec:results}

We now present our numerical results for the thermodynamic properties of finite SSH chains. All quantities are expressed in dimensionless form as defined in the previous section. We analyze the system with open boundary conditions allowing particle exchange  in Sec.~\ref{subsec:fixed_sites}, and with conserved particle number in Sec.~\ref{subsec:fixed_particles}.

\subsection{Fixed chain length: Open boundary conditions}
\label{subsec:fixed_sites}

We first consider SSH chains of fixed length $N$ in contact with a particle reservoir at chemical potential $\mu$. This scenario corresponds to the grand canonical ensemble and models experimental situations where charge carriers can be exchanged with the environment, such as in doped semiconductors or gated quantum systems.

\subsubsection{Heat capacity landscape}

Since the system exchanges particles with the reservoir in this scenario, the relevant specific heat plotted in the following figures is the heat capacity at constant chemical potential, $C_\mu = (\partial \bar{E}/\partial T)_\mu - \mu (\partial \bar{N}/\partial T)_\mu$. This ensures that the energy fluctuation correctly accounts for the chemical work done by the reservoir. Consequently, the mean particle number $\bar{N}$ is not fixed and varies with temperature.

\begin{figure*}[ht]
    \includegraphics[width=\textwidth]{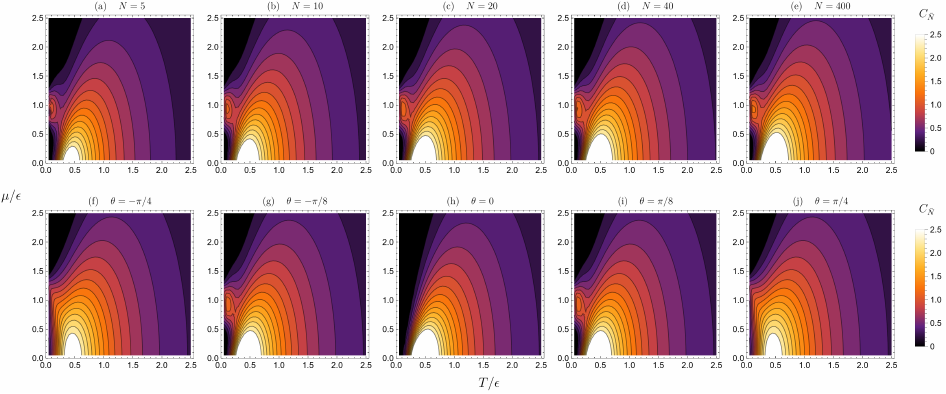}
    \caption{Heat capacity $C_{\bar{N}}/N$ landscape in the grand canonical ensemble. Here, $N$ represents the chain length (in unit cells), while $\bar{N}$ is the number of particles considered. Upper row (a)-(e): Evolution with the number of unit cells $N$ for a fixed hopping asymmetry $\theta=\pi/8$. Lower row (f)-(j): Evolution with the hopping angle $\theta$ for a fixed chain length of $N=20$ unit cells. All panels use a linear temperature scale starting at $T/\epsilon=0$. The double-peak structure is prominent at intermediate asymmetries when the chemical potential is comparable to the band scale ($\mu/\epsilon \sim 1$). In contrast, when $|\mu/\epsilon| \gg 1$ or in the fully dimerized limits ($\theta=\pm\pi/4$), the system enters a single activation regime and only one peak survives. Due to the $\theta \to -\theta$ symmetry of the spectrum, panels (f) and (j) are physically identical.}
    \label{fig:heat_capacity_2D}
\end{figure*}

Figure~\ref{fig:heat_capacity_2D} presents contour plots of the heat capacity $C_{\bar{N}}$ as a function of temperature $\tilde{T}$ and chemical potential $\tilde{\mu}$ for various system parameters. The upper row (panels (a)-(e)) shows the evolution with chain length for fixed hopping asymmetry $\theta = \pi/8$, while the lower row (panels (f)-(j)) displays different values of $\theta$ for fixed $N = 20$.

Several striking features emerge from these plots. For intermediate values of the hopping asymmetry ($\theta \neq 0, \pm\pi/4$), the heat capacity exhibits two local maxima separated by a local minimum at low temperatures ($\tilde{T} \lesssim 0.5$) and chemical potential $\tilde{\mu} \sim 1$. This local minimum is the valley between the two maxima of a Schottky-type double-peak anomaly; it is not a metastable phase and does not signal a phase transition. The two maxima mark the successive thermal activation of two groups of levels set by two different energy scales, and the minimum is simply the temperature window between them. The fact that the heat capacity is a second derivative of the thermodynamic potential does not by itself imply a transition, since a finite chain has an analytic free energy at all temperatures.
Furthermore, comparing panels (a)-(e), we observe that the anomaly becomes more pronounced as the chain length increases from $N=5$ to $N=400$:
the two maxima move apart and the intervening minimum deepens. This trend follows from the growth of the density of states and from the widening separation between the two energy scales as the chain lengthens, and should not be read as the onset of a sharp transition. For a one-dimensional non-interacting fermion chain with short-range hopping no finite-temperature transition exists in any limit, so what the sequence $N=5\to 400$ displays is the gradual sharpening of a crossover, not finite-size scaling toward criticality.
The lower row of Fig.~\ref{fig:heat_capacity_2D} demonstrates that the double-peak anomaly is highly sensitive to the parameter $\theta$. For the fully dimerized cases ($\theta = -\pi/4$ and $\theta = \pi/4$, panels (f) and (j)), only a single maximum appears, and the heat capacity monotonically decreases as $|\mu|$ increases.
A single peak is expected in these limits because a fully dimerized chain reduces to decoupled dimers with a single pair of levels $\pm\epsilon$, so only one energy scale is available and the heat capacity is a single two-level Schottky peak (see the analytic treatment below).
For the same reason panels (f) and (j) are essentially identical even though $\theta=-\pi/4$ and $\theta=+\pi/4$ differ: the spectrum in Eq.~(\ref{eq:energy_spectrum}) is invariant under $\theta\to-\theta$, so the two fully dimerized limits give the same set of energy magnitudes and hence the same heat capacity.
For the symmetric case $\theta = 0$ (panel (h), corresponding to $v=w$), a saddle point emerges instead of a clear minimum. Intermediate asymmetries (panels (g) and (i)) show well-developed double-peak anomalies. The second, low-temperature maximum appears only when the chemical potential is comparable to the band scale, $\tilde\mu\sim 1$: it is then that a group of levels sits close enough to $\mu$ to be thermally activated at low temperature and produce the first peak, while the rest of the band produces the second. When $|\tilde\mu|\gg 1$ or $|\tilde\mu|\ll 1$ the chemical potential leaves the system in a single activation regime and only one peak survives, which explains why the double-peak structure is confined to the region $\tilde\mu\sim 1$ in Fig.~\ref{fig:heat_capacity_2D}.
On the other hand, the heat capacity exhibits approximate symmetry under $\theta \to -\theta$, reflecting the equivalence of the two dimerization patterns in finite chains with collapsed wavefunctions at the boundaries. This is consistent with the symmetry of the energy spectrum in Eq.~(\ref{eq:energy_spectrum}).

\subsubsection{Temperature slices and density correlations}

\begin{figure*}
    \includegraphics[width=\textwidth]{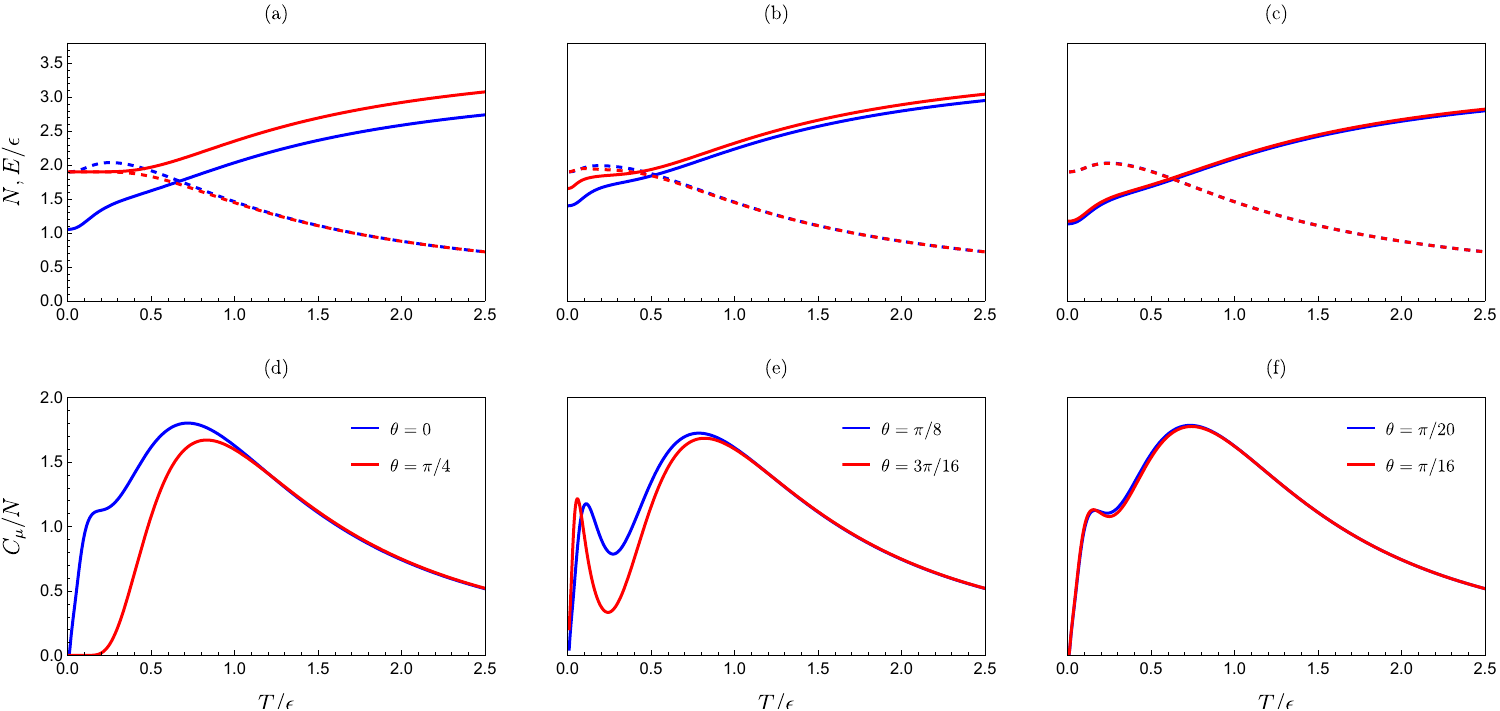}
    \caption{Temperature dependence of thermodynamic quantities in the grand canonical ensemble for a fixed chain length of $N=20$ unit cells and chemical potential $\mu/\epsilon=1.0$. Upper panel (a)-(c): Mean energy density $\bar{E}/\epsilon$ (solid lines) and mean particle number density $\bar{N}/N$ (dashed lines). Lower panel (d)-(f): Heat capacity per site $C_\mu/N$ evaluated at constant chemical potential. The horizontal axis uses a linear scale, confirming that the heat capacity properly vanishes as $T \to 0$ in accordance with the third law of thermodynamics.
%     \revnote{This figure is computed in the grand canonical ensemble at fixed $\mu$, so $\bar N$ varies with temperature (upper panel); state clearly in the caption and text that the heat capacity shown here is the fixed-$\mu$ heat capacity, not $C_{\bar N}$ at fixed $\bar N$, to resolve the apparent contradiction raised by the referee. Redraw the horizontal axis on a linear scale starting at $\bar T = 0$ for uniformity with Figs.~\ref{fig:heat_capacity_fixed_N} and \ref{fig:energy_fixed_N}, add subfigure labels, and check the low-temperature behavior: the heat capacity must vanish as $\bar T\to 0$ (third law), so any apparent rise near $\bar T=0$ is a logarithmic-axis artifact to be corrected.}
}
    \label{fig:slices}
\end{figure*}

Figure~\ref{fig:slices} provides a more detailed view by taking slices through the heat capacity landscape at fixed $\tilde{\mu} = 1.0$ and $N = 20$. The bottom panel shows $C_{\bar{N}}$ versus $\tilde{T}$ for different values of $\theta$, clearly illustrating the emergence of the second maximum at low temperature for non-dimerized cases [Fig. \ref{fig:slices}(e) and (f)].
Moreover, a saddle point behavior for the symmetric case $\theta = 0$, whereas
a single-peak structure arises for the dimerized case $\theta = \pi/4$, as shown in Fig. \ref{fig:slices}(d). Notice that
the increase in the height of the second maximum as $|\theta|$ increases from 0 indicates a more pronounced double-peak anomaly for larger hopping asymmetry. Moreover,
the top panel of Fig.~\ref{fig:slices} shows the corresponding energy density $\bar{E}/\epsilon$ (solid lines) and particle number density $\bar{N}/N$ (dashed lines). Several important observations can be made. For non-dimerized cases exhibiting the double-peak anomaly, there is a notable increase in $\bar{N}$ at temperatures corresponding to the local minimum in heat capacity. This suggests that the crossover region is characterized by enhanced particle density, consistent with increased accessibility of excited states. At the same time,  the energy density shows significant changes in slope at temperatures corresponding to heat capacity anomalies, confirming that these features are genuine crossover features of the density of states rather than numerical artifacts. Notice that at low temperatures, $\tilde{T} \lesssim 0.1$, all curves converge to similar values, while at high temperatures, $\tilde{T} \gtrsim 2$, the energy approaches $k_B T$ per degree of freedom, consistent with equipartition theorem expectations.

\subsubsection{Physical interpretation}

The double-peak anomaly revealed in the heat capacity can be understood in terms of the interplay between temperature, chemical potential, and the discrete energy spectrum. At low temperatures and moderate chemical potential, thermal excitations preferentially populate certain energy levels that are particularly sensitive to the hopping asymmetry. As temperature increases, additional levels become accessible, leading to a second regime of heat absorption (second maximum). The local minimum between these maxima represents a temperature range where the system has exhausted the readily accessible low-lying states but has not yet reached the energy scale for bulk excitations. This is precisely the structure of a two-scale Schottky crossover: writing the two characteristic scales as $\Delta_1$, the spacing between the levels nearest the chemical potential, and $\Delta_2\sim\epsilon(1-\cos 2\theta)^{1/2}$, the scale associated with the far edge of the band, the first maximum sits near $k_B T\sim\Delta_1$ and the second near $k_B T\sim\Delta_2$, with the minimum in the window $\Delta_1\lesssim k_B T\lesssim\Delta_2$. The separation $\Delta_2/\Delta_1$ grows with the hopping asymmetry and with the chain length, which is why the anomaly becomes more pronounced in both directions.

\begin{figure}
    \centering
    \includegraphics[width=0.5\linewidth]{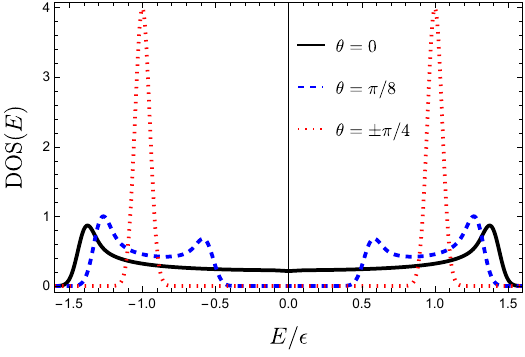}
    \caption{DOS for the values $N=20$ and various $\theta$-values.}
    \label{fig:DOS}
\end{figure}

This interpretation is supported by examining the density of states (DOS) implied by Eq.~(\ref{eq:energy_spectrum}), shown in Fig. \ref{fig:DOS}. The DOS explicitly reveals the origin of the Schottky-type thermal crossover. For the symmetric case ($\theta=0$), the gap is completely closed at $E=0$. However, for intermediate asymmetries such as $\theta=\pi/8$, the gap opens and van Hove singularities emerge at both the inner and outer band edges, producing four distinct peaks. The first heat capacity maximum corresponds to thermal excitations across the gap between the inner peaks, while the second maximum occurs when temperatures are high enough to explore the full bandwidth up to the outer peaks. In contrast, for fully dimerized systems ($\theta=\pm\pi/4$), the bands collapse into flat, isolated states at $\pm\epsilon$, resulting in two massive Dirac delta peaks and a single energy scale.

The fully dimerized limits admit a fully analytic check that makes the single-scale nature of the peak explicit. At $\theta=\pm\pi/4$ the chain reduces to decoupled two-level dimers with energies $\pm\epsilon$, so at complete filling the heat capacity per site takes the two-level Schottky form
\begin{equation}
\tilde C(\tilde  T) = 
%\left(\frac{\epsilon}{k_B T}\right)^{2}\,\frac{e^{\epsilon/k_B T}}{\left(1+e^{\epsilon/k_B T}\right)^{2}} =
\left(\frac{1}{\tilde T}\right)^{2}\frac{e^{1/\tilde T}}{\left(1+e^{1/\tilde T}\right)^{2}} ,
\label{eq:schottky}
\end{equation}
a single peak located near $\tilde T\simeq 0.42$, with no second maximum because only one energy scale is present. Please note that we are missing a factor of 4 in Eq. (\ref{eq:schottky}) arising from spin degeneracy and particle-hole symmetry. 
\begin{figure}
    \centering
    \includegraphics[width=0.5\linewidth]{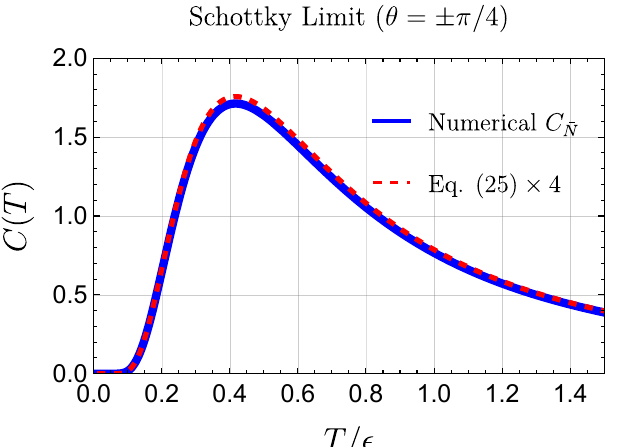}
    \caption{Overlay of the theoretical and numerical predictions of the two-level Schottky limit $\tilde{C}(T)$ for $\theta=\pm\pi/4$ and $N=20$.}
    \label{fig:schottky}
\end{figure}
Figure \ref{fig:schottky} shows an overlay of this theoretical limit and our numerical evaluation for $\theta=\pm\pi/4$ and $N=20$. The excellent quantitative agreement validates our numerical framework. The slight amplitude offset observed is not a numerical artifact, but a strict signature of finite-size physics. The hard-wall boundary conditions restrict the bulk standing waves to $N-1$ states. Consequently, the numerical specific heat per site intrinsically carries an exact $(N-1)/N$ scaling factor (yielding 0.95 for $N=20$) compared to the ideal thermodynamic limit assumed in the analytical equation. This factor precisely accounts for the amplitude reduction, demonstrating the robustness of our finite-size thermodynamic formulation.
Importantly, this thermodynamic crossover is fundamentally different from the topological phase transition at $\theta = 0$. While the topological transition is characterized by the closing of the bulk gap and the appearance/disappearance of protected edge states, namely, a boundary phenomenon, the thermodynamic crossover we observe reflects bulk properties encoded in the full energy spectrum. Indeed, the anomaly is most pronounced away from the topological critical point, indicating that the topological classification and the thermodynamic behavior are distinct.

\subsection{Fixed particle number: Canonical ensemble}
\label{subsec:fixed_particles}

We now consider the complementary scenario where the number of charge carriers is conserved ($\bar{N} = \text{const.}$) while the chain length and temperature vary. This corresponds to the canonical ensemble and models isolated systems or devices with controlled doping levels where particle exchange with the environment is suppressed.

\subsubsection{Heat capacity with constrained particle number}

\begin{figure*}[t]
    \includegraphics[width=\textwidth]{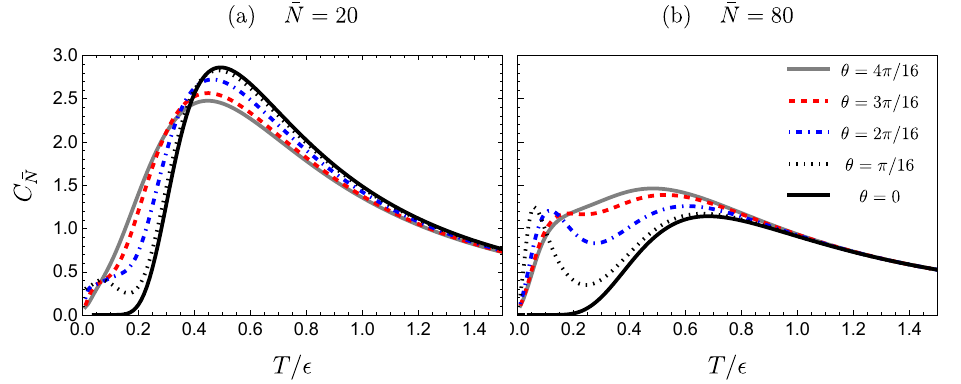}
    \caption{Heat capacity per site $C_{\bar{N}}/N$ in the canonical ensemble as a function of temperature for a strictly fixed mean number of particles. (a) Half-filling with $\bar{N}=20$. (b) Higher filling with $\bar{N}=80$. The number of unit cells is kept fixed at $N=40$ in both panels. A single-peak structure emerges at $\theta=0$ and at the fully dimerized limits, consistent with previous canonical studies at $\mu=0$ Ref.~[24, Fig.~2(b)]\cite{MunozArboleda2024}, while intermediate asymmetries develop a double-peak structure due to the interplay of two distinct energy scales. The curve legends in (b) stand for both figures.
  % \revnote{Add (a)/(b) labels to the two panels and label the individual $\theta$ curves. Include a direct comparison with the single-peak $C_{\bar N}$ reported for the SSH model at $\mu=0$ in Ref.~[24, Fig.~2(b)], and note that the double peak requires $\bar\mu\sim 1$, so the single peak at $\mu=0$ is the expected single-scale limit.}
   }
    \label{fig:heat_capacity_fixed_N}
\end{figure*}

Figure~\ref{fig:heat_capacity_fixed_N} shows the heat capacity $C_{\bar{N}}$ versus temperature for $N = 40$ sites and two different particle numbers: $\bar{N} = 20$ [half-filling, Fig.~\ref{fig:heat_capacity_fixed_N}(a)] and $\bar{N} = 80$ [higher filling, Fig.~\ref{fig:heat_capacity_fixed_N}(b)]. Several hopping asymmetries are compared.
Key observations are the following.  Even with fixed particle number, the double-peak anomaly --characterized by a local minimum in $C_{\bar{N}}$-- persists for non-dimerized cases. However, the effect is less pronounced than in the open boundary case, particularly at half-filling ($\bar{N} = 20$).
On the other hand, the anomaly becomes much more evident at higher filling factor, $\bar{N} = 80$. This suggests that particle number fluctuations, which are present in the grand canonical ensemble but suppressed here, play a role in amplifying the thermodynamic signatures. Despite of the quantitative differences, the qualitative behavior mirrors the open boundary case: fully dimerized systems show single peaks, the symmetric case exhibits saddle point behavior, and intermediate asymmetries display two-peak structure with an intervening minimum. Moreover,  the characteristic temperatures for heat capacity maxima and minima are comparable to those in the grand canonical ensemble, confirming that the underlying energy scales (set by $\epsilon$) govern the thermodynamic behavior in both ensembles.

\subsubsection{Energy density and chemical potential evolution}

\begin{figure*}    \includegraphics[width=1.\textwidth]{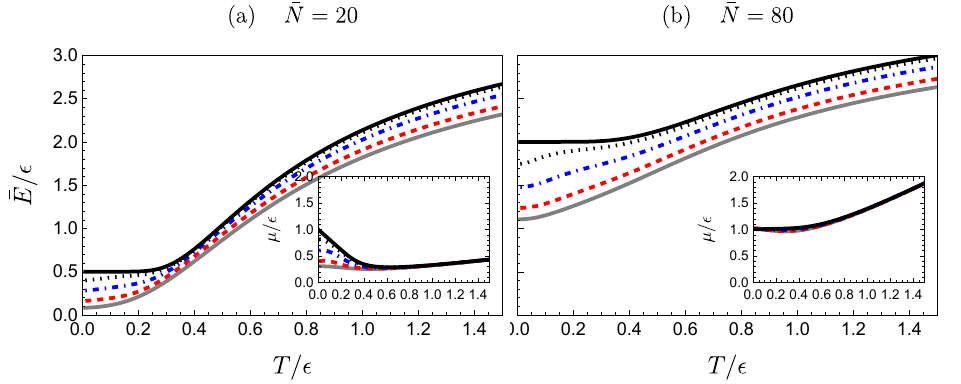}
    \caption{Energy density $\bar{E}/\epsilon$ as a function of temperature for different hopping asymmetries $\theta$ at a fixed mean number of particles. (a) $\bar{N}=20$. (b) $\bar{N}=80$. In both cases, the number of unit cells is fixed at $N=40$. The insets show the corresponding temperature dependence of the chemical potential $\mu/\epsilon$ required to strictly conserve the particle number $\bar{N}$ across the entire temperature range. The curve legends are the same as in Fig. \ref{fig:heat_capacity_fixed_N}(b).%\revnote{Add (a)/(b) panel labels and label the $\theta$ curves, for uniformity with the other figures.}
    }
    \label{fig:energy_fixed_N}
\end{figure*}

Figure~\ref{fig:energy_fixed_N} displays the energy density $\bar{E}/\epsilon$ and chemical potential $\tilde{\mu}$ (insets) as functions of temperature for the same systems as in Fig.~\ref{fig:heat_capacity_fixed_N}.
The energy density shows relatively weak dependence on the hopping parameter $\theta$ at fixed temperature, with differences becoming negligible at high temperatures where classical statistics dominate. The main effect of varying $\theta$ is to shift the low-temperature energy baseline, reflecting the change in ground state energy with dimerization.
More revealing is the behavior of the chemical potential, which decreases approximately linearly with temperature, with the slope and intercept depending strongly on $\theta$. This reflects the changing balance between entropy and energy as thermal excitations populate higher levels. At high-temperatures, $\tilde{T} \gtrsim 1$, all curves converge to a common approximately linear behavior, independent of $\theta$. This universality arises because at high temperatures, the system enters a classical regime where specific details of the hopping parameters become irrelevant. The transition between these two regimes occurs around $\tilde{T} \sim 1$, corresponding to the characteristic energy scale of the problem. Furthermore, at $\bar{N} = 80$, the chemical potential values are systematically higher than at $\bar{N} = 20$, as expected from Fermi statistics. The crossover to universal high-temperature behavior occurs at similar temperatures, indicating that the energy scale $\epsilon$ remains the controlling parameter.

\subsection{Comparison with related systems}

Our results can be contextualized within the broader literature on finite-size thermodynamics and topological systems. For example, in~\cite{MunozArboleda2024}, thermodynamics of the non-Hermitian SSH model was analyzed, showing that non-Hermiticity induces unconventional phase transitions related to exceptional points. Complementary thermodynamic approaches to topological systems were developed in~\cite{Kempkes2016UniversalitiesThermoTopo}, while thermodynamic and spectral properties of adiabatic Peierls chains were studied in~\cite{Weber2016AdiabaticPeierls}. In related SSH-like surface systems, metastable behavior associated with a grand canonical Peierls transition was reported in~\cite{Jeckelmann2015GrandCanonicalPeierls}. In contrast, our Hermitian system exhibits heat capacity anomalies and thermal crossovers driven by the interplay of dimerization, finite size, and thermal fluctuations, without requiring non-Hermitian physics. On the other hand, the heat capacity anomalies we observe bear resemblance to those in the quantum Ising chain~\cite{Pfeuty1970}, which also exhibits non-monotonic behavior near quantum critical points when finite-size effects are important. However, the SSH model lacks the order parameter and spontaneous symmetry breaking characteristic of the Ising transition.
Furthermore, studies of thermodynamic signatures of topological transitions~\cite{Ezawa2014} typically focus on fidelity susceptibility or entanglement entropy. Our work demonstrates that conventional thermodynamic quantities like heat capacity can also reveal rich crossover structure, albeit one that is distinct from the topological classification. We stress that double-peak heat capacities are not unusual: they are the generic fingerprint of two competing energy scales and appear as Schottky anomalies in a broad range of two-level and multi-level systems~\cite{pashchenko2002magnetic,custers2012destruction, hejtmanek2014magnetism,Appelgate,KURUMAJI2023169475}. %\revnote{Add references to Schottky double-peak anomalies in real two-level and multi-level materials (for example paramagnetic salts and rare-earth compounds); include only sources that can be verified.}
Our double-peak anomaly should also be compared directly with the single-peak $C_{\bar N}$ obtained for the SSH model at $\mu=0$ in Ref.~\cite{MunozArboleda2024}: the two results are consistent once it is recognized that the second peak requires a chemical potential comparable to the band scale, so the $\mu=0$ case sits in the single-scale regime where only one peak survives. 
In our finite SSH chain, the anomaly similarly reflects a purely microscopic mechanism—the existence of a finite set of accessible low-energy states separated by distinct gaps—rather than spontaneous symmetry breaking or collective critical behavior.
Finally, the growth of the anomaly with system size [Fig.~\ref{fig:heat_capacity_2D}, panels (a)-(e)] reflects the sharpening of a crossover as the density of states increases, not finite-size scaling toward a critical point; since this free-fermion chain has no finite-temperature transition, no critical exponents are associated with it.

\subsection{Experimental implications}

The thermodynamic signatures we have identified could be experimentally accessible in several platforms. As a first example, the SSH model has been implemented in LC circuit arrays~\cite{Ningyuan2015,Lee2018,PerezPedraza2024,ANZUREZ2025417609}, where the hopping parameters correspond to coupling capacitances. Heat capacity measurements could be realized through calorimetry or by monitoring the circuit temperature-dependent admittance.
Ultracold atoms in optical lattices provide exquisite control over system parameters~\cite{Atala2013}. Thermodynamic quantities can be extracted from time-of-flight measurements and density profiles. The double-peak crossover might manifest as anomalous relaxation dynamics.
Moreover,  while photons are bosons rather than fermions, SSH-like physics has been realized in photonic lattices~\cite{Li2014SSH}. Thermodynamic analogues involving classical field fluctuations could potentially exhibit similar crossover structure.
Furthermore, recent realizations of 1D topological insulators in germanene nanoribbons~\cite{Klaassen2025} suggest that finite SSH chains might be fabricated at nanometer scales. Thermal transport and thermoelectric measurements in such systems could reveal the predicted heat capacity anomalies.

\section{Conclusions}
\label{sec:conclusions}

In this work, we have presented a comprehensive thermodynamic analysis of finite Su-Schrieffer-Heeger chains, revealing rich crossover structure beyond the well-known topological classification. Our main findings and their implications can be summarized as follows. We have identified a double-peak anomaly in the heat capacity of non-dimerized SSH chains, characterized by a local minimum flanked by two maxima at low temperatures and moderate chemical potential. This feature is a Schottky-type thermal crossover between two energy scales, not a phase transition: the free energy of a finite chain is analytic in temperature, and a one-dimensional non-interacting fermion chain with short-range hopping has no finite-temperature transition in any limit. The anomaly becomes more pronounced as the hopping asymmetry increases (away from the symmetric point $v=w$) and as the chain length grows, a trend that reflects the widening separation of the two energy scales and the growth of the density of states rather than emergent criticality.
Our results demonstrate that the topological classification and the thermodynamic behavior are distinct. While the topological transition at $\theta=0$ is characterized by edge state properties and winding numbers, the thermodynamic crossovers we observe are bulk effects encoded in the full density of states. The anomaly is most evident precisely away from the topological critical point, indicating that systems in the topologically trivial or non-trivial phases can both exhibit complex thermodynamic behavior.
We have shown that the thermodynamic signatures are sensitive to system size, with the anomaly deepening and widening as the number of unit cells increases. This is the gradual sharpening of a crossover, not finite-size scaling toward a critical point, and even modest system sizes ($N \sim 20$) exhibit clear signatures, making experimental detection feasible in current platforms.
By comparing grand canonical  and canonical  ensembles, we have demonstrated that particle number fluctuations play an important role in shaping thermodynamic behavior. The anomaly persists in both ensembles but is more pronounced when particle exchange is permitted, indicating that coupling to a particle reservoir enhances thermodynamic signatures. Finally, the hopping parameter ratio $v/w$ (or equivalently $\theta$) provides an experimentally accessible control parameter for tuning the system through different thermodynamic phases. This opens possibilities for engineering thermal properties in nanoscale devices and quantum simulators based on SSH physics.

Several avenues for future research emerge from this work. A systematic study of the system size dependence of the two peak positions, heights, and widths, together with the temperatures of the two maxima and the intervening minimum, would characterize the two energy scales that govern the crossover and their dependence on $\theta$ and $N$. Since the finite free-fermion chain hosts no finite-temperature transition, this analysis targets the crossover scales rather than critical exponents or a universality class.
Moreover, our equilibrium thermodynamic analysis could be extended to study relaxation dynamics, particularly in the crossover region. Quench dynamics starting from non-equilibrium initial states might reveal interesting crossover behavior between different thermal regimes.
The inclusion of electron-electron interactions, e.g., nearest-neighbor Coulomb repulsion, would transform the SSH model into an interacting topological system. Interactions could compete with or enhance the crossover structure we have identified, potentially leading to exotic phases such as topological Mott insulators~\cite{Guo2013SSH}. The interacting and Su-Schrieffer-Heeger-Hubbard variants of the model, whose ground-state phase diagrams and analytic structure have been studied in detail~\cite{Ahmadi2020,FengXing2022,CaiLiYao2022,XingFeng2023,InteractingSSH2024,SSHHK2025}, provide a natural setting for such an extension. Extension to two-dimensional topological insulators like Haldane or Kane-Mele models~\cite{Haldane1988,Kane2005} would test the generality of our findings. The interplay between topological invariant Chern numbers or Z$_2$ indices and thermodynamic crossovers in 2D systems remains largely unexplored. Furthermore, the calculation of the entanglement entropy and its temperature dependence could provide additional insight into the nature of the double-peak anomaly. A detailed proposals for measuring heat capacity in specific experimental platforms, including topoelectrical circuits, cold atoms, photonic systems, would facilitate experimental verification of our predictions. Challenges include isolating the system sufficiently for thermodynamic measurements while maintaining contact with temperature and particle reservoirs. Moreover, the generalization to SSH models with beyond-nearest-neighbor and long-range hopping~\cite{Li2014SSH,LiMiroshnichenko2019,PerezGonzalez2019} could reveal how the crossover structure depends on the range of the couplings. Long-range hopping modifies the topological properties; its effect on thermodynamics remains to be explored. Finally, real experimental systems inevitably contain disorder in hopping amplitudes or on-site energies. Understanding how the double-peak anomaly and heat capacity features survive disorder would be important for experimental detection.

In conclusion, our work demonstrates that finite topological systems possess thermodynamic complexity beyond their topological classification. The double-peak thermal crossovers and heat capacity anomalies we have identified represent experimentally accessible signatures that could be observed in quantum simulator platforms. More broadly, our results highlight the importance of considering thermodynamic properties alongside topological invariants for a complete understanding of topological materials, particularly in the finite-size regime relevant to nanoscale devices and quantum technologies.

\section*{Acknowledgments}

A.R. acknowledges financial support from CIC-UMSNH México under grant 18371. C.V. acknowledges support from ANID (Chile) through Fondecyt Regular grant No. 1250206. J.C.P.P. acknowledges support from SECIHTI under the program “Estancias Posdoctorales por México”
with CVU number 671687.  We acknowledge ``El Faro'' for guiding us through the darkest moments of this research.%We thank [colleagues] for useful discussions.

%\revnote{The following ten references, requested by the referees and verified as genuine, must be added to \texttt{SSH\_thermodynamics\_references.bib}. Their BibTeX entries are provided in the commented block below (cite keys: SirkerBoundary2014, Shin1997Tridiagonal, LiMiroshnichenko2019, PerezGonzalez2019, Ahmadi2020, FengXing2022, CaiLiYao2022, XingFeng2023, InteractingSSH2024, SSHHK2025).}

\bibliographystyle{unsrtnat}
\bibliography{SSH_thermodynamics_references}% Produces the 

\end{document}